\documentclass[10pt]{iopart}
\usepackage[utf8]{inputenc}
\usepackage{color}
\usepackage[T1]{fontenc}
\usepackage{graphicx}

\expandafter\let\csname equation*\endcsname\relax
\expandafter\let\csname endequation*\endcsname\relax
\usepackage{amsmath}
\usepackage{amsfonts}
\usepackage{braket}
\usepackage{hyperref}
\usepackage{bm}
\expandafter\let\csname eqalign\endcsname\relax
\usepackage{nccmath}

\usepackage{txfonts}
\usepackage{eurosym}

\newcommand{\diag}{\text{diag}}
\newcommand{\id}{\mathbb{I}}
\newcommand{\dcrit}{D_{\text{crit}}}
\newcommand{\lambdacrit}{\lambda_{\text{crit}}}
\newcommand{\dtrk}{D_\text{\tiny{TRK}}}
\newcommand{\trk}{\text{\tiny{TRK}}}
\newcommand{\hamiltonian}{\mathcal{H}}
\newcommand{\rhozero}{\rho^{\text{\tiny(0)}}}
\newcommand{\rhotrk}{\rho^{\text{\tiny(TRK)}}}
\newcommand{\fihomodyne}{F^\text{(hd)}}

\newcommand{\changed}[1]{{#1}}
\newcommand{\TTchanged}[1]{{#1}}

\begin{document}
\title{Probing the diamagnetic term in light-matter interaction}
\author{Matteo A. C. Rossi}
\address{Quantum Technology Lab,
Dipartimento di Fisica, Universit\`a degli Studi di Milano,
20133 Milano, Italy}
\address{School of Mathematical Sciences, The University of Nottingham, Nottingham NG7 2RD, United Kingdom}
\author{Matteo Bina}
\address{Quantum Technology Lab,
Dipartimento di Fisica, Universit\`a degli Studi di Milano,
20133 Milano, Italy}
\author{Matteo G. A. Paris}
\address{Quantum Technology Lab, Dipartimento di Fisica,
Universit\`a degli Studi di Milano, 20133 Milano, Italy}
\address{INFN, Sezione di Milano, I-20133 Milano, Italy}
\author{Marco G. Genoni}
\address{Quantum Technology Lab, Dipartimento di Fisica,
Universit\`a degli Studi di Milano, 20133 Milano, Italy}
\address{Department of Physics and Astronomy, University College London,
London WC1E 6BT, United Kingdom}
\author{Gerardo Adesso}
\address{School of Mathematical Sciences, The University of Nottingham, Nottingham NG7 2RD, United Kingdom}
\author{Tommaso Tufarelli}
\address{School of Mathematical Sciences, The University of Nottingham, Nottingham NG7 2RD, United Kingdom}

\date{\today}
\begin{abstract}
\TTchanged{We address the quantum estimation of the diamagnetic, or $A^2$, term in an effective model of light-matter interaction featuring two coupled oscillators. First, we calculate the quantum Fisher information of the diamagnetic parameter in the interacting ground state. Then, we find that typical measurements on the transverse radiation field, such as homodyne detection or photon counting, permit to estimate the diamagnetic coupling constant with near-optimal efficiency in a wide range of model parameters. Should the model admit a critical point, we also find that both measurements would become asymptotically optimal in its vicinity. Finally, we discuss binary discrimination strategies between the two most debated hypotheses involving the diamagnetic term in circuit QED. While we adopt a terminology appropriate to the Coulomb gauge, our results are also relevant for the electric dipole gauge. In that case, our calculations would describe the estimation of the so-called transverse $P^2$ term. The derived metrological benchmarks are general and relevant to any implementation of the model, cavity and circuit QED being two relevant examples.}
\end{abstract}

\maketitle
\noindent\paragraph{Introduction.} The ultra-strong coupling (USC) regime of light-matter interaction has recently attracted much interest, thanks to impressive advances in theory and experiments \cite{Gunter2009,Maissen2014}. A number of interesting phenomena, of fundamental physical appeal and potential impact on future quantum technologies, arise in this regime, including the formation of a nontrivial ground state \cite{Ciuti2005}, dynamical Casimir-like effects \cite{DeLiberato2007}, and the possibility of quantum phase transitions \cite{Nataf2010,Emary2004}.

Loosely speaking, the USC regime is entered when the light-matter coupling constant, say $\lambda$, is a non-negligible fraction of the bare frequencies of light and matter. In this situation the rotating-wave approximation breaks down, and one cannot neglect the so-called ``counter-rotating terms" in the interaction Hamiltonian \cite{Ciuti2005,Cohen1997}. While this brings about exciting new physics, it results also in formidable computational challenges. Effective models featuring a small number of degrees-of-freedom (modes), such as the Dicke Hamiltonian \cite{Nataf2010}, are extremely useful in this context: they can reveal crucial features of the new regime while keeping computations tractable. 

\TTchanged{However, there is disagreement in the literature regarding the specific form that these few-degrees-of-freedom models should take, in particular concerning the presence (or otherwise) of the so-called diamagnetic or $A^2$ term. We recall that such a term originates from the general form of the Coulomb gauge minimal coupling Hamiltonian
\begin{equation}
\hamiltonian_{\text{min}}=\sum_{j}\frac{(\hat{\mathbf{p}}_j-q_j\hat{\mathbf{A}})^2}{2 m_j}+V(\hat{\mathbf{x}}_1,\hat{\mathbf{x}}_2,...)+\hamiltonian_{\text{EM}},\label{Hmin}
\end{equation}
which describes the interaction of the quantised electromagnetic field with non-relativistic charged particles. In the above equation $m_j$, $q_j$, $\hat{\mathbf{x}}_j$ and $\hat{\mathbf{p}}_j$ indicate respectively the mass, charge, position and canonical momentum of the $j-$th particle, $\hat{\mathbf{A}}$ is the vector potential operator, $V$ is the instantaneous Coulomb potential energy, while $\hamiltonian_{\text{EM}}$ is the free Hamiltonian of the transverse radiation field \cite{Cohen1997}. $\hamiltonian_{\text{min}}$ clearly contains a contribution proportional to the squared vector potential $\hat{\mathbf{A}}^2$, i.e. the diamagnetic term.

In cavity QED (and other similar set-ups) many authors derive their effective models by a direct few-mode truncation of Hamiltonian \eqref{Hmin}, thus explicitly retaining an $A^2-$like term affecting the radiation modes of interest. The importance of such term is well recognized in preventing the Dicke phase transition \cite{Birula1979,Raza1991,Bamba2014,Chirolli2012} and in limiting the effectively achievable light-matter coupling \cite{DeLiberato2014}. Within such framework, the coupling strength associated with the $A^2$ term obeys precise constraints (see below), whose origin can be traced back to the Thomas-Reiche-Kuhn (TRK) sum rule \cite{Nataf2010}.}

In contrast, other authors hold that the few-mode approximation is better carried out in the electric dipole gauge, that is, they apply a canonical transformation to Hamiltonian~\eqref{Hmin} \textit{before} the derivation of an approximate model. This yields an effective Hamiltonian without diamagnetic term, and seemingly capable of a phase transition \cite{Vukics2014,Vukics2012,Keeling2007}. \TTchanged{Yet, it has been argued that such derivations neglect the contribution of the so-called transverse $P^2$ term, which in the electric dipole gauge plays a role mathematically analogue to $A^2$ \cite{Bamba2014}}.
 
In the context of circuit QED, Ref.~\cite{Nataf2010} maintains that the diamagnetic term can be avoided by appropriately tuning the circuit parameters, a claim which has generated further debate \cite{Viehmann2011,Ciuti2012,Viehmann2012,Jaako2016}. In fact, in circuit QED, there is even disagreement on the most appropriate microscopic description underlying the Dicke model. It is indeed not clear whether the most reliable starting point for the derivation of an effective model should be a minimal coupling Hamiltonian \cite{Viehmann2012} or a direct quantization of macroscopic circuit variables \cite{Jaako2016}.

\TTchanged{In this Letter we take a complementary approach to the $A^2$-term debate: instead of trying to address the question theoretically, by deriving a Dicke model from first principles, we propose that it may be settled experimentally via an appropriate measurement. We note that steps in this direction have been taken in recent theoretical proposals \cite{Garcia-Ripoll2015,Lolli2015}. Moreover, present-day experimental results seem to support the thesis of Nataf and Ciuti~\cite{Nataf2010}, i.e. that the $A^2$ term may be neglected in some circuit QED systems \cite{Yoshihara2016}. In this work we contribute to these efforts by exploiting the tools of quantum metrology. We address the estimation of the $A^2$ coupling constant via measurements on the coupled ground state, thus deriving a theoretical benchmark relevant to any experimental implementation of the model (circuit QED and cavity QED being two relevant special cases). Our work is a first step towards setting precise metrological standards for any experiment probing the diamagnetic parameter.}

\begin{figure}[t]
\centering
	\includegraphics{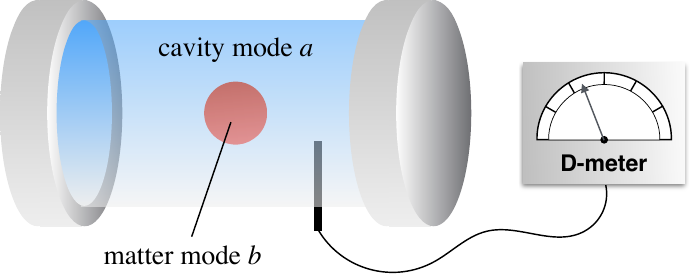}
	\caption{The parameter $D$, quantifying the strength the diamagnetic term in the Dicke Hamiltonian \eref{eq:hamiltonian}, may be estimated experimentally by performing an appropriate measurement on the coupled light-matter system. \changed{Here we find the ultimate bound on the information about $D$ that can be extracted with such a measurement}. Our results also suggest that realistic measurements such as homodyne detection or photon counting of the cavity field $a$ are well suited to this task in a broad range of the model parameters.\label{fig:dmeter}}
\end{figure}

We consider a general Dicke model of two coupled oscillators featuring
counter-rotating terms and an $A^2$-like term with unknown coupling
constant $D$. As the establishment of a nontrivial ground state is
perhaps the simplest signature of the USC, we find it natural to ask how
much information about $D$ is contained in this state (i.e., how sensitive the state is to variations in $D$), and how efficiently such information can be extracted by typical measurements (see Fig.~\ref{fig:dmeter}
for an illustration). Exploiting the tools of quantum estimation theory
\cite{Braunstein94,lnp649,amari2007methods,Escher2011,BrodyHughston1998,
Paris2009}, we are able to give quantitative answers to these
questions.

Besides deriving the ultimate bounds on the estimation precision of $D$,
we find that typical measurements on the transverse cavity field, such as homodyne detection or photon counting, allow for the precise estimation of the diamagnetic term in a wide range of model parameters. Namely, homodyne detection is optimal when the coupling constant $\lambda$ is vanishing, still providing a close-to-optimal performance for $\lambda\lesssim20\%$ of the bare mode frequencies. This regime of
relatively small coupling (yet within the USC) is relevant in both
cavity QED \cite{Anappara2009} and circuit QED experiments
\cite{Niemczyk2010,Forn-Diaz2010}, and is well described within the two-mode approximation \cite{Tufarelli2015}.
\TTchanged{Moreover, if the Dicke model of interest describes a collection of $N$ two-level emitters, this is also a regime in which low-lying excitations are well described by a coupled oscillator model, even for relatively small values of $N$: for example, this is the case for a spin-2 Dicke model ($N=5$) \cite{Tufarelli2015}.}

While we tackle the problem in its full generality, we also obtain compact analytical results for three
important limiting cases: (i)  $D\simeq 0$; (ii) $D \simeq \dtrk$, where
$\dtrk$ is the value derived from the TRK sum rule \cite{Nataf2010,Tufarelli2015}; (iii) $D\simeq \dcrit$, where $\dcrit$
is the threshold for the Dicke phase transition. In case the model
admits a phase transition, we further find that the considered
measurements become approximately optimal near the critical point. However we point out that finite-size effects, not included in our model, may become increasingly relevant as criticality is approached \cite{Rodriguez-Lara2010,Bina2016,Wang2014}.

We complement our analysis by discussing the problem of binary
discrimination \cite{HelstromBook,bergou10} between the choices $D=0$ and $D=\dtrk $, which may be relevant in circuit QED \cite{Viehmann2011,Nataf2010}. 
We then discuss the experimental feasibility of the proposed measurements, and conclude the Letter by providing an outlook of further research questions stemming from our work.

Before proceeding we note that in this Letter we adopt a language appropriate to the Coulomb gauge, such that field operators describe excitations of the transverse (and gauge invariant) radiation field, while what we call ``matter mode" in fact describes a combination of matter and field properties \cite{Cohen1997}. In this form, our results can be used to determine the most appropriate coupling constant of the diamagnetic term. \TTchanged{In \ref{dipollo} we show that our results can be readily adapted to describe the estimation of the $P^2$ term in dipole gauge \cite{Bamba2014}.}

\paragraph{The model.}
Let us start by introducing the Hamiltonian of the system. We denote with $a$ the photonic cavity mode, of bare frequency $\omega_a$, and with $b$ the matter mode, with frequency $\omega_b$. After setting $\hbar = 1$, the Hamiltonian reads
\begin{equation}\label{eq:hamiltonian}
	\hamiltonian = \omega_a a^\dagger a + \omega_b b^\dagger b + \lambda (a + a^\dagger)(b + b^\dagger) + D(a+a^\dagger)^2,
\end{equation}
where $\lambda$ is the light-matter coupling strength and the last is the diamagnetic term, with the constant $D$ being the object of our discussion.

Being quadratic in the field operators, $\hamiltonian$ is easily diagonalized and can be written, up to an irrelevant constant, as 
	$\hamiltonian = \omega_U p_U^\dagger p_U + \omega_L p_L^\dagger p_L,$
where $p_{U,L}$ are bosonic operators for the upper and lower polaritonic modes. The eigenfrequencies $\omega_{U,L}$ are reported in \cite{Tufarelli2015} and in \ref{sec:covariance_matrix_of_the_ground_state}.

If the diamagnetic term is absent, i.e. we set $D=0$ in Eq.~\eref{eq:hamiltonian}, we recover the standard Dicke Hamiltonian, with a quantum phase transition at $\lambda = \lambdacrit \equiv \sqrt{\omega_a \omega_b}/2$. 
 On the other hand, if the diamagnetic term is included, it must satisfy the inequality
 \begin{equation}\label{eq:condition}
 	D\ge \dcrit \equiv \lambda^2/\omega_b - \omega_a/4
 \end{equation}
 in order for the Hamiltonian in Eq.~\eqref{eq:hamiltonian} to be bounded from below. If $D$ assumes the `TRK value' \cite{Tufarelli2015},
 \begin{equation}\label{eq:dtrk}
 D =\dtrk\equiv {\lambda^2}/{\omega_b},
 \end{equation}
 then Eq.~\eref{eq:condition} is always satisfied and the phase transition is suppressed.

 The ground state of the Hamiltonian is the vacuum state of the polaritonic modes, i.e.~a Gaussian state with zero mean and covariance matrix $\sigma_0 = \id / 2$. Expressing this state in terms of the initial modes $a$ and $b$, we obtain a non-trivial, two-mode squeezed Gaussian state with covariance matrix $\sigma = S \sigma_0 S^T$, where the suitable symplectic transformation $S$ is reported in \ref{sec:covariance_matrix_of_the_ground_state}.

\paragraph{Quantum estimation methods.} Let us now address the problem of estimating $D$, and in doing so we shall review the basics of quantum estimation theory. In order to determine the value of the unknown parameter, one collects measurement outcomes from experiments and builds an estimator $\hat D$, i.e.~a function of the data that will return our best guess for the value of $D$. Assuming that the estimator is unbiased, its precision, in the limit of a large number $n$ of measurements, is bounded from below by the Cram\'er-Rao bound \cite{Cramer1946} $\text{Var} \hat D \geq [n F(D)]^{-1}$, where  $F(D)$ is the Fisher information (FI) of the probability distribution $p(x|D)$, defined as $F(D) = \int dx p(x|D) [\partial_D \ln p(x|D)]^2$. Here $p(x|D)$ is the conditional probability that the outcome of the measurement is $x$, if the parameter value is $D$. If we define the quantum Fisher information (QFI) $H(D)$ as the supremum of the FI over all possible quantum measurements described by positive operator-valued measures (POVMs), we obtain the quantum Cram\'er-Rao bound, i.e.~the ultimate lower bound on the achievable precision in the estimation of $D$ \cite{Paris2009}.

The QFI with respect to $D$ can be evaluated analytically for Gaussian states \cite{Monras2013,Jiang2014,Bina2016} and for the case at hand is given by \begin{equation}\label{eq:qfi}
	H(D) = \Tr[\Omega^T\partial\sigma \Omega \Phi],
\end{equation}
where $\partial$ denotes the derivative with respect to $D$, $\Omega=\oplus_{i=1}^2\binom{\ 0 \ \ 1}{-1\ 0}$ is the symplectic matrix, and the matrix $\Phi$ satisfies the linear equation
		$\partial\sigma= 2\sigma \Omega \Phi \Omega^T \sigma - \frac12 \Phi.$
The expression for the QFI is quite cumbersome and we report it in \ref{sec:qfi}. In the following we analyse its behavior in the cases of interest for our discussion. We also note that the QFI is a homogeneous function of the parameters of the Hamiltonian: $H(\alpha D, \alpha \omega_a,\alpha \omega_b,\alpha \lambda) = \alpha^{-2} H(D,\omega_a,\omega_b,\lambda)$.

The QFI for the two-mode ground state is a relevant benchmark for the precision of any measurement on the system. In realistic experiments, however, it will only be possible to measure a limited number of system observables. For definiteness, we shall focus here on typical measurements that can be performed on the transverse radiation field $a$. It is thus interesting to compute also the QFI $H_a(D)$ for the reduced state of mode $a$, which is a squeezed thermal state, see \ref{sec:qfi}.

Homodyne detection corresponds to the measurement of the quadrature operator $\hat x(\phi) = (a e^{-i \phi} + a ^\dagger e^{i \phi})/\sqrt{2}$, where $\phi$ is an arbitrary angle. For the reduced state $\rho_a$, the probability distribution for the outcome of $x(\phi)$ is a normal distribution with zero mean and covariance $\sigma_{11} \cos ^2\phi + \sigma_{22} \sin ^2\phi$; the corresponding FI, $\fihomodyne$ is reported in  \ref{sec:qfi}. It is further argued that the FI has a maximum at  $\phi = 0$: this has the important consequence that the best measurement angle does not depend on the (yet unknown) $D$ parameter. On the other hand, the FI for photon counting cannot be determined analytically as it involves an infinite sum of the terms $p(n) = \braket{n|\rho|n}$, i.e.~the probabilities of finding a photon in the Fock state $\ket n$ \cite{Marian2012,Bina2016}; we evaluate it numerically in a truncated Fock space.

\paragraph{Results: Probing the diamagnetic term.}
The main results of this paper, namely the analytical expressions for the QFI and FI for the homodyne detection (reported in Appendix B) are valid for any values of the model parameters, provided the Hamiltonian \eqref{eq:hamiltonian} admits a ground state. In what follows, we focus on the parameter regimes that are most relevant from a theoretical and from an experimental point of view.

Let us start by focusing on the regime of relatively small coupling within the USC. The QFI has a non vanishing limiting value
\begin{equation}
	H(D) ={2}{(4 D + \omega_a)^{-2}} +O (\lambda^2) \:,
\end{equation}
showing how, in this regime, smaller values of $D$ can be estimated more efficiently, yet a large number of measurements is needed to  neutralise the contribution of $\omega_a$ and obtain a precise estimation. Remarkably, the FI for the homodyne measurement saturates $H(D)$ up to second order in $\lambda$, indicating that, no matter what the true value of $D$ is, we can estimate it with optimal efficiency by detecting the quadrature of mode $a$ when $\lambda$ is sufficiently small.
In detail, if we consider the plausible scenario $D\lesssim \dtrk$, the ratio between FI and $H(D)$ is
\begin{equation}\label{eq:ratio_small_lambda_D0}
	\frac{\fihomodyne(D\lesssim \dtrk)}{H(D\lesssim \dtrk)} =  1-\frac{8 \omega_a^2}{(\omega_a+\omega_b)^4}\lambda ^2 + O(\lambda^4).
\end{equation}

\begin{figure}\centering
	\includegraphics{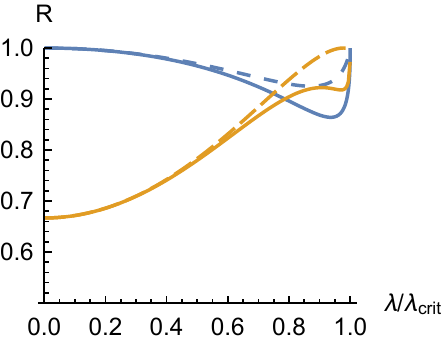}\includegraphics{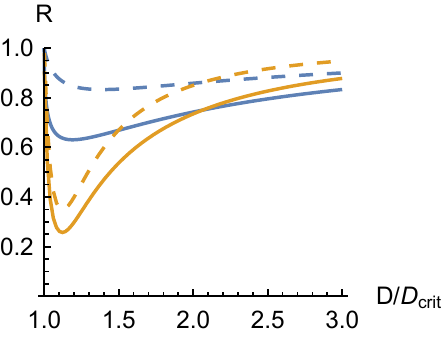}
	\caption{Left: Ratio $R$ between the FI and the QFI for the two modes (solid) and the photon mode (dashed) for $D = 0$ as functions of $\lambda$. Homodyne detection is highlighted in blue and photon counting in orange. The plot is obtained for $\omega_a = \omega_b = 1$.  Homodyne detection is optimal for small $\lambda$ and both measurements saturate the QFI when $\lambda \approx \lambdacrit$.  Right: Plot of the same quantities when $D$ gets close to the critical value $\dcrit$, for $\lambda = 1$ and $\omega_a = \omega_b = 1$. Both measurements saturate the QFI in the limit $D \rightarrow \dcrit$ and $D/\dcrit \rightarrow \infty$, with homodyne detection reaching generally higher ratios in the region close to the critical value.}
	\label{fig:hd_pc_near_dcrit_and_D0}
\end{figure}

\begin{figure}\centering
	\includegraphics{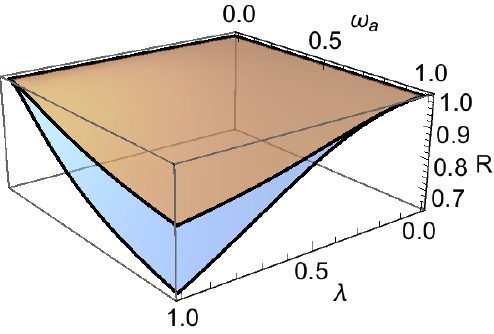} \includegraphics{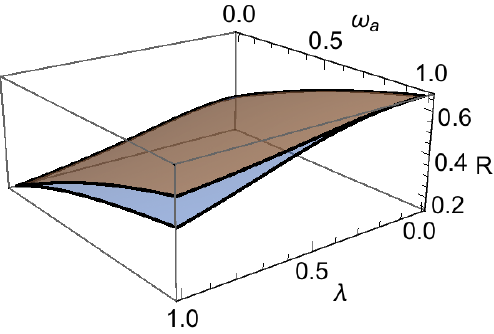}
\caption{Ratios $\fihomodyne(D_\trk)/H(D_\trk)$ (blue) and $\fihomodyne(D_\trk)/H_a(D_\trk)$ (orange) for homodyne detection (left) and photon counting (right), at $D=\dtrk$ and  $\omega_b = 1$. Homodyne detection is optimal for small $\lambda$, \changed{but allows to extract a relevant amount of information (about $70\%$), also when $\lambda \simeq \omega_{a,b}$.} Photon counting is optimal for large $\omega_a$ and small $\lambda$.}
	\label{fig:trk_ratio}
\end{figure}

We can now investigate the estimation properties of the diamagnetic parameter $D$ without constraining the value of the coupling constant $\lambda$ and thus, when possible, also at the critical point. For the case $D=0$ the phase transition appears when the coupling reaches the critical value $\lambdacrit = \sqrt{\omega_a\omega_b}/2$.  In this case the QFI for $D$ diverges as
\begin{equation}
	H(D=0) \sim  {\omega_b}/[{8\omega_a(\lambda -\lambdacrit )^{2}}] \:,
\end{equation}
and the homodyne and the photon counting measurements both saturate the QFI in this limit, as shown in Fig.~\ref{fig:hd_pc_near_dcrit_and_D0} (left panel).
If $D$ approaches the TRK value $D\simeq\dtrk$ [cf. Eq.~\eref{eq:dtrk}], the phase transition is suppressed and there is no criticality in the system. The QFI for the two-mode state (for any $\lambda$) reads
\begin{equation}
	H\left(\dtrk\right)=\frac{2}{\omega _a^2}-\frac{16 \lambda ^2 \omega _a \omega _b}{\left[4 \lambda ^2 \omega _a +\omega _b \left(\omega _a+\omega _b\right){}^2\right]^2}.
\end{equation}
$H(\dtrk)$ is clearly higher for small frequencies of the radiation mode, whereas it decreases in the case of high frequencies.
With respect to $\lambda$, $H(\dtrk)$ has a minimum at $\lambda^2 = {\omega_b (\omega_a+\omega_b)^2}/{4\omega_a}$ and the asymptotic value $2/\omega_a^2$ for large and small $\lambda$.
The ratio between the FI for the two measurements and the QFI is shown in Fig.~\ref{fig:trk_ratio} as a function of $\lambda$ and $\omega_a$. One can also verify that the expansion in Eq.~\eref{eq:ratio_small_lambda_D0} is in fact a good guideline for remarkably large values of $\lambda$, implying that homodyne would be an excellent measurement strategy in most experimentally relevant situations. For example, setting $D=\dtrk$ and $\omega_a=\omega_b=2\lambda$, we still get a ratio of $81/92 \simeq 0.88$, while the approximation of Eq.~\eref{eq:ratio_small_lambda_D0} would yield $7/8\simeq 0.87$.

In general, in the vicinity of the critical point $\dcrit$, the asymptotic behavior of the two-mode QFI is
\begin{equation}
	H(D) \underset{D \rightarrow \dcrit}{\sim} \frac{1}{8\,(D-\dcrit)^{2}},
\end{equation}
that is, the QFI diverges, meaning that the amount of information that can be extracted per measurement increases enormously. This is another confirmation of the role of quantum criticality in the enhancement of estimation \cite{Bina2016,Zanardi2007,Zanardi2008}. We point out, however, that we neglect finite-size effects which may become relevant near the critical point. Nonetheless, it is reasonable to believe that these effects translate into a smoothing of the QFI, for either the atomic ensemble or the radiation mode (see Ref. [33] for the finite-size case without the diamagnetic term).

The QFI for the reduced state of one mode saturates the two-mode QFI $H(D)$ when the system is close to the critical point. This is remarkable: optimal estimation of $D$ around criticality can be achieved by measuring only a part of the system. Moreover, it can effectively be achieved with feasible experiments such as homodyne detection or photon counting (see Fig.~\ref{fig:hd_pc_near_dcrit_and_D0}). The FI for the $x$-quadrature measurement indeed saturates the QFI as $D$ gets close to the critical value $\dcrit$,
\begin{equation}
	\frac{\fihomodyne(D)}{H(D)} \sim 1-\frac{16 \lambda ^2 \omega _a^{3/2}}{4 \lambda ^2 \omega _a \omega _b+\omega _b^4}(D-\dcrit)^{1/2}.
\end{equation}
We further find numerically that the photon counting measurements can also saturate the QFI near the critical point.

\paragraph{Results: Quantum discrimination.} 
\label{par:discrimination}
We finally tackle the problem of discriminating between the two values
of $D$ that are most argued in the recent circuit-QED literature: $D=0$ and $\dtrk$. By
making a measurement on a subsystem, typically the cavity field, we want
to decide which of the two hypotheses is correct. Quantum mechanics
poses a limit to the discriminating ability, quantified by the Helstrom
bound \cite{HelstromBook,bergou10}: if we label $\rhozero$ and $\rhotrk$, respectively, the ground state in the two hypotheses, the lowest probability of error in the discrimination between these two states  is $P_e = \frac 12 \left[1- {\mathbb{D}}(\rhozero,\rhotrk)\right]$,
where ${\mathbb{D}}$ denotes the trace distance. The Helstrom bound is based on the assumption that the system has the same probability to be in either of the two states, reflecting the lack of any \emph{a priori} information on which hypothesis is the correct one. The Helstrom bound calculated on the two-mode state can only be used as an ideal benchmark, as it is saturated by a (typically unfeasible) collective measurement on the two modes.
A more practical benchmark is given by the Helstrom bound calculated on the reduced state of  mode $a$, i.e.~in the discrimination between $\rhozero_a$ and $\rhotrk_a$. Being these states mixed, the trace distance cannot be computed analytically, and a diagonalization of the density operators expanded on a truncated Fock basis is required. The ensuing bound on the probability of error, $P_e^{(a)}$, is less stringent than the two-mode bound.

\begin{figure}
	\centering
	\includegraphics{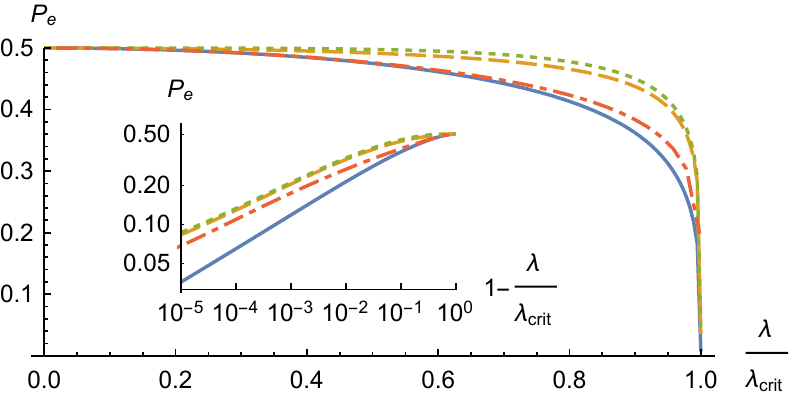}
	\caption{The probabilities of error $P_e$, given by the Helstrom bound on the two-mode system (solid blue), $P_e^{(a)}$, for the photon mode (dot-dashed orange), $P_{e}^{\text{(hd)}}$ for homodyne detection (dashed yellow) and $P_e^\text{(pc)}$ for photon counting (dotted green) as functions of $\lambda/\lambdacrit$, with $\omega_a = \omega_b = 1$. In the inset, a log-log plot of the same quantities as $\lambda$ approaches $\lambdacrit$. The probability of error is close to $1/2$ for small $\lambda$, vanishing when $\lambda \rightarrow \lambdacrit$, i.e.~when, for $D=0$, the system is close to the quantum phase transition. The Helstrom bound on the mode $a$ behaves differently from the two-mode bound near the critical value. The two proposed discrimination schemes, although not saturating the single-mode bound, have the same behavior in the limit $\lambda \rightarrow \lambdacrit$.}
	\label{fig:discrimination}
\end{figure}

We show the dependence of $P_e$ and $P_e^{(a)}$ on $\lambda/\lambda_c$ in Fig.~\ref{fig:discrimination}: the probability of error is close to $1/2$ for small $\lambda$ and vanishes when $\lambda \rightarrow \lambda_c = \sqrt{\omega_a \omega_b}/2$ in the case $D = 0$. The dependence on $\lambda$ in the neighborhood of $\lambda_c$ is $P_e \sim (\lambda_c - \lambda)^{1/4}$. For $P_e^{(a)}$ a numerical fit shows an exponent around $1/5$. Thus we find that the discrimination between the two hypotheses is hard in the regime of small coupling, while criticality allows for a great improvement, should the optimal strategy be found.

To this end, we checked the performance of two feasible discrimination strategies using either the $\hat x(0)$ quadrature measurement or photon counting on the mode $a$. The former exploits the fact that the probability distribution of the outcome of the quadrature measurement is a Gaussian with variance $\sigma_{11}$. In the case $D =0$, when $\lambda \rightarrow \lambdacrit$, $\sigma_{11} \sim (\lambdacrit-\lambda)^{-1/2}$. On the other hand, for $D = \lambda^2/\omega_b$, $\sigma_{11}$ does not depart much from its limit for $\lambda \rightarrow 0$, which is $1/2$.
Thus, close to the critical value, the Gaussian distribution for $D= \lambda^2/\omega_b$ is very narrow compared to the distribution for $D=0$.

We can thus set up a readily feasible discrimination experiment as follows: if the outcome of the experiment is $|x| < 2\sigma_{11}^{(\trk)}$ \footnote{The choice of $2\sigma_{11}^{(\trk)}$ here is arbitrary. The optimal threshold depends on the parameters of the problem. Even optimizing over this parameter we do not saturate the Helstrom bound.} then the state of the system is $\rhotrk$, otherwise it is $\rhozero$. The corresponding probability of error is
\begin{equation}
P_{e}^{\text{(hd)}}  = \int_{0}^{2\sigma_{11}^{(\trk)} }\mathcal{N}(0,\sigma^{(0)}_{11}) dx + \int_{2\sigma_{11}^{(\trk)}}^\infty\mathcal{N}(0,\sigma^{(\trk)}_{11}) dx
\end{equation}
where the first (second) term is the probability of detecting $\rhozero$ ($\rhotrk$) when the actual state was $\rhotrk$ ($\rhozero$), and $\mathcal{N}(\mu,\sigma)$ is the normal distribution with mean $\mu$ and variance $\sigma$.


In a photon counting experiment, we can exploit the fact that, when $D = 0$, the average photon number diverges near the phase transition \cite{Bina2016}, while it is close to zero if the phase transition is suppressed by the presence of the $A^2$ term. Thus, we can discriminate between the two hypotheses by setting a threshold photon number $n_T$ and assigning any outcome below that threshold to the hypothesis $\rhotrk$ and any outcome above to $\rhozero$. The probability of error in this case would be
\begin{equation}
P_e^\text{(pc)} = \frac 12 \left(\sum_{n=0}^{n_T}p_{0}(n) + 1 - \sum_{n=0}^{n_T}p_{\trk}(n) \right).
\end{equation}
We find that the optimal threshold value is $n_T=0$, i.e.~the discrimination is between no photons and any number of photons. This is a harder question to settle in a realistic experiment.

The resulting $P_{e}^{\text{(hd)}}$ and $P_e^\text{(pc)}$ are compared to the Helstrom bounds in Fig.~\ref{fig:discrimination}. We see that the homodyne scheme is slightly better than the photon counting one. Interestingly, these two feasible discrimination schemes have the same behavior as the Helstrom bound $P_e^{(a)}$ on the reduced state when approaching the critical point, although neither of them is optimal.

\begin{figure}
		\centering
		\includegraphics[width=7cm]{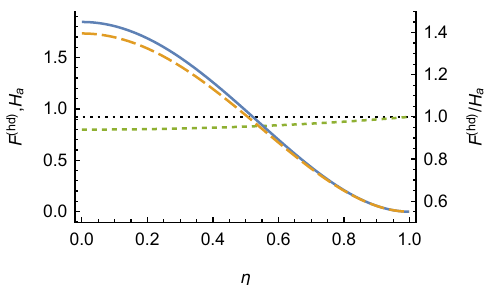}
		\caption{Effect of experimental imperfections on the estimability of $D$. Due to non-ideal conditions in the procedure of cavity field extraction, we assume that the cavity mode $a$ is affected by a pure-loss channel before the measurement. We plot the single-mode QFI of the field mode (solid blue), FI for homodyne detection (dashed orange), in units of the inverse-square frequency, and their ratio (dotted green, right scale) as functions of the loss parameter $\eta$, for $\omega_a = \omega_b = 1$ and $\lambda = 0.2$, assuming that $D = \lambda^2 / \omega_b$. The estimability of $D$ inevitably decreases with $\eta$, but the ratio between the FI and the QFI does not decrease and tends to one (dotted black line). This suggests that homodyne detection may be a robust measurement for the estimation of $D$ in realistic experimental conditions.}
		\label{fig:lossy_homodyne}
	\end{figure}

\paragraph{Measuring the cavity field.} Before closing, we would like to comment on the feasibility of the proposed measurements, which require access to the cavity mode $a$ (see Fig.~\ref{fig:dmeter}). One possible way to access the intra-cavity field is to suddenly switch off the coupling $\lambda$, a drastic yet experimentally feasible procedure \cite{Gunter2009}. Allowing one of the cavity mirrors to have a small but finite transmissivity, one may subsequently collect the cavity output field (i.e.~the radiation that gradually leaks out into the external world). In absence of light-matter coupling and other losses, the cavity output field can be used to extract the full quantum statistics of mode $a$ just before the switch-off \cite{Gardiner2004}. A crucial open question, however, is how coherent the switching process can be, i.e., how well it can preserve the quantum state of light. Importantly, our results can be generalized in future work to take into account experimental imperfections in the switching process: due to finite transients, losses and other decoherence mechanisms, one would end up measuring a deteriorated version of the original cavity field $a$.  It is reasonable to assume that the whole process could be modelled as a quantum channel acting on the reduced state of the cavity mode, just before the latter is measured. Insofar as these imperfections can be described by a Gaussian channel, the problem could be attacked by a straightforward generalization of the tools employed here. Importantly, the relevant noise parameters must be known in advance for this approach to work within the paradigm of single-parameter estimation. As an example, we illustrate the special case of a Gaussian pure-loss channel \cite{Serafini2005}, described by a loss parameter $\eta \in [0,1]$ quantifying the probability of single-photon loss (with $\eta = 1$ meaning complete loss, i.e. the output state is always the vacuum). The effect on the estimability of $D$ is shown in Fig. \ref{fig:lossy_homodyne}. The QFI and FI inevitably decrease with $\eta$, but the ratio does not, signifying that homodyne detection may be a robust measurement to estimate the diamagnetic term in presence of decoherence.
	
\paragraph{Outlook and conclusions.}
We investigated the detection of the diamagnetic term in a Dicke model of light-matter interaction, formulating the problem in terms of quantum parameter estimation or quantum state discrimination. We obtained the ultimate quantum limits to the rate at which information can be extracted from the ground state of the coupled system, and we discussed the performance of two typical measurements on the cavity field, homodyne and photon counting, that allow for efficient estimation of the parameter of interest. This efficiency becomes optimal in experimentally relevant regimes, such as the small-coupling regime \changed{in which the coupling is up to $20\%$ of the mode frequencies}. 

To conclude, we would like to indicate some possible directions for future work. By employing multiparameter quantum estimation theory \cite{Paris2009}, our study can be expanded to cover more realistic situations in which other parameters of the model, e.g. loss rates or the coupling $\lambda$, are not known exactly. The role of finite-size effects could also be explored, in particular near criticality where a coupled-oscillator model might be inaccurate \cite{Rodriguez-Lara2010} (this, however, will require techniques beyond the Gaussian formalism employed here). The estimability of the $A^2$-term from a thermal state of the coupled system could also be addressed: in this scenario radiation will be continuously emitted by the cavity, without the need to have a fast modulation of the coupling constant.

Finally, we remark that the ideas presented in this Letter may be generalised to test the validity of alternative or more complex models of light-matter interaction; for example, alternative Dicke models for circuit QED such as that proposed in \cite{Jaako2016}, or models with additional terms to describe electrostatic contributions (e.g. dipole-dipole interactions between the atoms), or effective contributions from higher harmonics of the cavity field \cite{Tufarelli2015}.

\ack
This work has been supported by EU through the Erasmus+ Placement
programme, the Marie Sk\l odowska-Curie Action H2020-MSCA-IF-2015
(project ConAQuMe) and the collaborative Project QuProCS
(Grant Agreement 641277), the EPSRC UK (grant No.~EP/K026267/1),
the European Research Council (ERC StG GQCOP, Grant No.~637352),
the Royal Society (Grant No.~IE150570), and by UniMI through the
H2020 Transition Grant 15-6-3008000-625.

\appendix
\section{Covariance matrix of the ground state} 
\label{sec:covariance_matrix_of_the_ground_state}
As is shown in \cite{Tufarelli2015}, the USC Hamiltonian in Eq. \eref{eq:hamiltonian} can be cast to the diagonal form of Eq. (2), where the frequencies $\omega_U$, $\omega_L$ are
\begin{equation}\label{eq:eigenfrequencies} 
	\omega_{U,L}^2 = \frac{\omega_a^2 + 4 D \omega_a + \omega_b^2}2
	 \pm \sqrt{ \left(\frac{\omega_a^2 + 4 D \omega_a - \omega_b^2}2\right)^2 + 4 \lambda^2\omega_a\omega_b}.
\end{equation}
The symplectic matrix that transforms the vector $\left(a,b,a^\dagger,b^\dagger\right)$ of the original-mode creation and annihilation operators into the vector of polaritonic modes $\left(p_U,p_L,p_U^\dagger,p_L^\dagger\right)$, is given by

\begin{equation}
	\bar S = \begin{pmatrix}
 \cos \theta\; f_+ \left(\mfrac{\omega_U}{\omega_a}\right) & -\sin \theta\; f_+ \left(\mfrac{\omega_U}{\omega_b}\right) & \cos \theta\; f_- \left(\mfrac{\omega_U}{\omega_a}\right) & -\sin \theta\; f_- \left(\mfrac{\omega_U}{\omega_b}\right) \\
 \sin \theta\; f_+ \left(\mfrac{\omega_L}{\omega_a}\right) & +\cos \theta\; f_+ \left(\mfrac{\omega_L}{\omega_b}\right) & \sin \theta\; f_- \left(\mfrac{\omega_L}{\omega_a}\right) & \cos \theta\; f_- \left(\mfrac{\omega_L}{\omega_b}\right) \\
 \cos \theta\; f_- \left(\mfrac{\omega_U}{\omega_a}\right) & -\sin \theta\; f_- \left(\mfrac{\omega_U}{\omega_b}\right) & \cos \theta\; f_+ \left(\mfrac{\omega_U}{\omega_a}\right) & -\sin \theta\; f_+ \left(\mfrac{\omega_U}{\omega_b}\right) \\
 \sin \theta\; f_- \left(\mfrac{\omega_L}{\omega_a}\right) & \cos \theta\; f_- \left(\mfrac{\omega_L}{\omega_b}\right) & \sin \theta\; f_+ \left(\mfrac{\omega_L}{\omega_a}\right) & +\cos \theta\; f_+ \left(\mfrac{\omega_L}{\omega_b}\right) \\
\end{pmatrix}
\end{equation}
where
\begin{equation}
	\cos 2\theta = \frac{\omega_a^2+4 D \omega_a-\omega_b^2}{\omega_U^2-\omega_L^2} \qquad \sin (2 \theta )=-\frac{4 \lambda  \sqrt{\omega_a \omega_b}}{\omega_U^2-\omega_L^2}.
\end{equation}
and we define
\begin{equation}
	f_\pm(x) = \frac{1}{{2}}\left(\sqrt{x}\pm \frac {1}{\sqrt{x}}\right).
\end{equation}
The ground state of the diagonalized Hamiltonian is the vacuum, i.e. a Gaussian state with no displacement and a covariance matrix $\sigma_0 = \id / 2$ in the basis of the quadratures $x_{U,L} = (p_{U,L}+ p_{U,L}^\dagger)/\sqrt{2}$, $y_{U,L} = -i(p_{U,L}-p_{U,L}^\dagger)/\sqrt{2}$.

To obtain the covariance matrix $\sigma$ in the original modes $a$ and $b$, transform $\sigma_0$ using a symplectic matrix $S$that can easily be obtained from $\bar S$. After some simplifications, we have

\begin{equation}
	\sigma =  S \sigma_0  S^T =  \left(
		\begin{array}{cccc}
		 \frac{\omega_a \left(\omega_b^2+\omega_L \omega_U\right)}{\omega_L \omega_U (\omega_L+\omega_U)} & 0 & -\frac{2 \lambda  \omega_a \omega_b}{\omega_L \omega_U (\omega_L+\omega_U)} & 0 \\
		 0 & \frac{2 \left(\omega_a^2+4 D \omega_a+\omega_L \omega_U\right)}{2 \omega_a \omega_L+2 \omega_a \omega_U} & 0 & \frac{2 \lambda }{\omega_L+\omega_U} \\
		 -\frac{2 \lambda  \omega_a \omega_b}{\omega_L \omega_U (\omega_L+\omega_U)} & 0 & \frac{\omega_b \left(\omega_a^2+4 D \omega_a+\omega_L \omega_U\right)}{\omega_L \omega_U (\omega_L+\omega_U)} & 0 \\
		 0 & \frac{2 \lambda }{\omega_L+\omega_U} & 0 & \frac{\omega_b^2+\omega_L \omega_U}{\omega_b (\omega_L+\omega_U)} \\
		\end{array}
		\right)
\end{equation}

\section{Quantum Fisher information and Fisher information for homodyne detection and photon counting} 
\label{sec:qfi}
\subsection{QFI for the two-mode state} 
\label{sub:qfi_for_the_two_mode_state}

Given $\sigma$, we can calculate the quantum Fisher information (QFI) using Eq. \eqref{eq:qfi}.
If the state is pure, the solution of Eq. (6) is $\Phi = - \partial \sigma$. This in turn yields the following expression for the QFI:
\begin{equation}
H(D) = 2 (\partial\sigma_{1,1} \partial\sigma_{2,2}+2\partial\sigma_{2,4} \partial\sigma_{3,1} + \partial\sigma_{3,3} \partial\sigma_{4,4}).
\end{equation}
Here we report the general expression of the QFI (after some manipulations)
	\begin{equation}\label{eq:qfi_explicit}
	\begin{split}
		H =  \frac{2 \omega_a^2 \omega_b^2}{\omega_L^4 \omega_U^4 (\omega_L+\omega_U)^4} & \left\{ (\omega_b^2 \left[16 D^2 \omega_a^2+8 D \omega_a \left(\omega_a^2+3 \omega_b^2+2 \omega_L \omega_U \right) \right. \right. \\
		 &  \left. \quad + \omega_a^4+4 \omega_L \omega_U \left(\omega_a^2+\omega_b^2\right)+6 \omega_a^2 \omega_b^2+\omega_b^4 \right] \\
		& \left. -8 \lambda ^2 \omega_a \omega_b \left(4 D \omega_a+\omega_a^2+2 \omega_b^2\right)+32 \lambda ^4 \omega_a^2 \right\}.
\end{split}
	\end{equation}
The relevant cases are discussed in the paper.
\subsection{QFI for the reduced state of mode $a$} 
\label{sub:qfi_for_the_reduced_state}

The covariance matrix $\sigma_a$ of the reduced state $\rho_a$ of the photonic mode is simply the upper left diagonal block of $\sigma$. The corresponding state is a squeezed thermal state
\begin{equation}
	\rho_a = S(r) \nu_{th} S^\dagger(r) = \frac 1 {1+N} \sum_m^\infty \left(\frac{N}{1+N}\right)^m S(r) \ket{m}\bra{m} S^\dagger(r),
\end{equation}
where $N$ is the number of thermal photons and $r$ is the squeezing parameter:
\begin{align}
	N & = \frac{1}{2} \left(\sqrt{\frac{\left(\omega _b^2+\omega _L \omega _U\right) \left(4 D \omega _a+\omega _a^2+\omega _L \omega _U\right)}{\omega _L \omega _U \left(\omega _L+\omega _U\right){}^2}}-1\right) \\
	r & = \frac{1}{4} \log \frac{\omega _L \omega _U \left(4 D \omega _a+\omega _a^2+\omega _L \omega _U\right)}{\omega _a^2 (\omega _b^2+\omega _L \omega _U)}.
\end{align}
The QFI for the reduce state can be obtained easily by solving Eq. (6) of the paper for $\Phi$. Being $\sigma_a$ diagonal, we simply find that $\Phi$ is diagonal with
\begin{eqnarray}
	\Phi_{11} &= 2\frac{(\partial \sigma_a)_{11}+4(\sigma_a)_{11}^2(\partial \sigma_a)_{22}}{16 (\sigma_a)_{11}^2 (\sigma_a)_{22}^2-1}\\
	\Phi_{22} &= 2\frac{(\partial \sigma_a)_{22}+4(\sigma_a)_{22}^2(\partial \sigma_a)_{11}}{16 (\sigma_a)_{11}^2 (\sigma_a)_{22}^2-1}.
\end{eqnarray}
The corresponding QFI is 
\begin{equation}
	H_a = \frac{4 \left(2 \sigma_{11}^2 \left(\partial \sigma_{22}\right)^2+2 \sigma_{22}^2 \left(\partial\sigma_{11}\right)^2+\partial\sigma_{11}\partial\sigma_{22}\right)}{16 \sigma_{11}^2 \sigma_{22}^2-1}.
\end{equation}

\subsection{Fisher information for the homodyne detection} 
\label{sub:fi_for_the_homodyne_detection}
In a homodyne detection experiment one can measure the field mode quadrature at an arbitrary phase, i.e. the expected value of the operator $\hat x(\phi) = {(a e^{-i \phi} + a ^\dagger e^{i \phi})}/{\sqrt{2}}$.

The probability density for the outcome of a homodyne measurement is easily obtained from the Wigner function of the reduced state of the photon mode $\rho_a = \Tr_b[\rho]$, which is a Gaussian distribution with zero mean and covariance matrix $\diag( \sigma_{11}, \sigma_{22})$.

The probability $p_\phi(x)$ of the outcoume $x(\phi)$ for the measurement is the marginal distribution of the Wigner function, obtained after a rotation of an angle $\phi$ in the phase space:
\begin{equation}
	\begin{split}
	p_\phi(x) & = \int dy W[\rho_a](x \cos \phi - y \sin \phi, x \sin \phi + y \cos \phi) \\
	 & = \mathcal{N}(0,\sigma_{11} \cos ^2\phi + \sigma_{22} \sin ^2\phi),
	\end{split}
\end{equation}
that is, a normal distribution with variance $\sigma_{11} \cos ^2\phi + \sigma_{22} \sin ^2\phi$.
The Fisher information (FI) for this distribution is easily obtained to be
\begin{equation}
	F_{\phi}(D) = \int dx \frac{[\partial p_\phi(x)]^2}{p_\phi(x)} =\frac{ \left(\partial\sigma_{11} \cos ^2\phi + \partial\sigma_{22} \sin ^2\phi \right)^2}{2 \left(\sigma_{11} \cos ^2\phi+\sigma_{22}\sin ^2\phi\right)^2}.
\end{equation}

It is easy to check that the FI $F_{\phi}(D)$ has maxima for $\phi = 0, \pi/2$, in which
\begin{equation}
	F_0(D) = \frac{(\partial\sigma_{11})^2}{2\sigma_{11}}, \qquad F_{\frac\pi2}(D) = \frac{(\partial\sigma_{22})^2}{2\sigma_{22}}.
\end{equation}
Numerical analysis indicates that $F_0(D) \geq F_{\frac\pi2}(D)$, thus $\phi = 0$ is the optimal angle to perform the homodyne measurement.
\TTchanged{\section{Dicke models in the dipole gauge}\label{dipollo}
Our calculations can be readily adapted to the study of Dicke models in the electric dipole gauge. In such a case, our Hamiltonian may be written as 
\begin{equation}\label{eq:hamiltonian2}
\hamiltonian_{\text{dip}} = \bar\omega_a a^\dagger a + \bar\omega_b b^\dagger b +  \bar\lambda(a + a^\dagger)(b + b^\dagger) + \bar D(b+b^\dagger)^2,
\end{equation}
where $\bar\omega_a,\bar\omega_b$, $\bar{\lambda}$, $\bar D$ are the bare frequencies and light-matter coupling constant relevant to the new gauge. Note how the coupling parameter $\bar D$ is now associated with the $P^2$ term \cite{Bamba2014}. In the dipole gauge, the operators $b, b^\dagger$ describe physical degrees of freedom of matter, thanks to the equivalence between canonical and kinetic momentum. On the other hand, the field operators $a,a^\dagger$ no longer describe the transverse radiation field, but are ``contaminated" by matter properties \cite{Cohen1997}. From Eq.~\eqref{eq:hamiltonian2} it is evident that the results presented in our manuscript can be easily translated in the dipole gauge, by simply swapping the role of $a$ and $b$ in all calculations. Then, the parameter to be probed becomes the $P^2$ coupling constant $\bar D$. Specifically, our calculations relative to homodyne detection and photon counting indicate that efficient estimation of the $P^2$ term is achievable through measurements on the matter degree of freedom.}
\section*{References}
\bibliographystyle{iopart-num}
\bibliography{aqquadro}
\end{document}